\documentclass[groupedaddress]{revtex4}

\usepackage{graphicx}
\usepackage{amssymb}
\usepackage{amsmath}

\begin{document}

\title{The ``Yin-Yang Grid'': An Overset Grid in Spherical Geometry}

\author{Akira Kageyama}
\email{kage@jamstec.go.jp}
\author{Tetsuya Sato}
\affiliation{Earth Simulator Center, 
Japan Agency for Marine-Earth Science and Technology,
Yokohama 236-0001, Japan}

\begin{abstract}
A new kind of overset grid, named Yin-Yang grid, 
for spherical geometry is proposed.
The Yin-Yang grid is composed of two identical
component grids that are combined
in a complemental way to cover a spherical surface
with partial overlap on their boundaries.
Each component grid is a low latitude part
of the latitude-longitude grid.
Therefore the grid spacing is quasi-uniform
and the metric tensors are simple and analytically known.
One can directly apply 
mathematical and numerical resources that have been written
in the spherical polar coordinates or latitude-longitude grid.
The complemental combination of the
two identical component grids enables us to 
make efficient and concise programs.
Simulation codes for geodynamo and mantle convection simulations
using finite difference scheme based on the Yin-Yang grid are
developed and tested.
The Yin-Yang grid is suitable for massively parallel computers.
\end{abstract}


\maketitle

\section{Introduction}
Since the Earth is composed of spherical layers,
computer simulations of the Earth's interior,
such as geodynamo and 
mantle convection simulations,
need efficient spatial discretization schemes
in spherical shell geometry.
The spectral method \citep{glatzmaier:1984} 
has been the major tool in the geodynamo simulation;
all six 
codes \citep{dormy:1998,christensen:1999,
sakuraba:1999,tilgner:1999,takahashi:2001} in the
benchmark test in \citet{christensen:2001}
and other codes \citep[e.g.,][]{kuang:1999,ishihara:2002}
use
the spherical harmonics expansion method in the horizontal space.
However, 
the importance of non-spectral (or point-based) 
approaches in the dynamo simulation is 
now increasingly recognized
to simulate more realistic geodynamo regime with 
smaller Ekman numbers \citep{chan:2001}.
The pursuit of point-based approaches started earlier 
in the mantle convection simulations, because the mantle's intense
spatial variation of viscosity and the phase transitions
makes the spectral approach not fit to the problem.
Although the spectral method for the mantle convection 
prospered in 1980s and 90s
\citep{machetel:1986,glatzmaier:1988,
bercovici:1989,zhang:1995,harder:1996},
the finite element method is rapidly growing in this field
\citep{baumgardner:1985,bunge:1995,
zhong:2000,tabata:2000,richards:2001}.
There are also a couple of codes that uses 
the finite element method in the geodynamo simulation
\citep{chan:2001,matsui:2002}.
The finite difference or finite volume method 
is applied for the mantle convection by
\citet{ratcliff:1996,iwase:1996,hernlund:2003}.
The finite difference method has been used
for the core convection and
the geodynamo simulation by the authors from 1990s
\citep{kageyama:1993,kageyama:1995,kageyama:1997,
kageyama:1997b,kageyama:1997c,kageyama:1999,
ochi:1999,li:2002},
in which the latitude-longitude grids
in the spherical polar coordinates
is used with radius $r$ ($r_i\le r\le r_o$),
colatitude $\theta$ ($0\le \theta\le \pi$),
and longitude $\phi$ ($0\le \phi < 2\pi$).
Since the finite difference method enables us
to make highly optimized programs for
massively parallel computers,
especially massively parallel vector supercomputers
like the Earth Simulator \citep{habata:2003},
we further exploit the possibility of the finite difference method
for simulations in spherical shell geometry
by improving the base grid system.

It is known that the latitude-longitude grid
has two numerical problems;
the coordinate singularity
and the grid convergence near the poles.
Since the coordinate singularity is not a real singularity
(the pole is not singular point of physical functions),
one can solve the basic equations on the poles
by applying the l'Hospital's rule
on the pole grids \citep[e.g.,][]{kageyama:1995}.
The computational cost for this pole grid solver is negligible.

The problem of the grid convergence is more serious.
In order to relax the severe restriction on the time step,
one has to apply a filter so that
the grid spacing on the sphere becomes
\textit{effectively} quasi-uniform.
The amount of information abandoned by the filter is
estimated by the number of grid points that are
\textit{effectively} present and that \textit{actually} present
in the computational space;
suppose one has a latitude-longitude grid of a spherical surface
of unit radius with inter mesh angles
$\Delta$ in both colatitude ($\theta$)
and longitude ($\phi$).
The azimuthal grid spacing, which is $\Delta$ in the equator,
converges in higher latitudes.
When a filter enables an effectively 
quasi-uniform grid with spacing $\Delta$
on the sphere,
the number of effective grid points
is estimated by 
$(\int_0^{\pi} \sin\theta d\theta \int_0^{2\pi} d\phi)
/ \Delta^2 = 4\pi/\Delta^2$.
While the number of actual grid points
in the computational space is given by
  $(\int_0^{\pi} d\theta \times \int_0^{2\pi} d\phi) / \Delta^2 
  = 2\pi^2/\Delta^2$.
Therefore sizeable ratio
of information, $(2\pi^2-4\pi)/2\pi^2 \sim 36\%$ 
of the latitude-longitude grid, is abandoned in vain 
by the filtering
at each simulation step.
In addition to this computational inefficiency,
the filter has non-negligible computational costs.
In our geodynamo simulation code using latitude-longitude grid,
in which a Fast Fourier Transform (FFT)-based filtering 
procedure is applied,
the filter routine can take more than $30\%$ of the total 
execution time.

Note that the above problem of the grid redundancy 
in the latitude-longitude grid 
comes only from the region of high latitudes.
The remaining part of the latitude-longitude 
grid---the low latitude region--has 
rather desirable feature for numerical simulations;
it is an orthogonal grid, it has simple metric tensors,
and it has quasi-uniform grid spacings.
This observation leads us to the idea of 
a new spherical grid proposed in this paper.

Since there is no grid mesh that is orthogonal all over the
spherical surface and, at the same time, free of 
coordinate singularity or grid convergence,
we decompose the spherical surface into subregions.
The decomposition, or dissection, enables us to cover each subregion
by a grid system
that is individually orthogonal and singularity-free.
This \textit{divide-and-rule} approach has been used with good success
in the computational aerodynamics that incorporates complex geometry
of aircraft's body with wings/stores/blades.

The dissection of the computational domain generates
internal border or boundary  between the subregions.
There are two different approaches to handle the 
internal boundaries.
One is the patched grid method \citep{rai:1986}
and the other is the
overset grid method \citep{chesshire:1990}.
In the patched grid approach, the subdomains contact one another
without any overlap on their borders.
In the overset grid method,
on the other hand,
the subdomains partially overlap one another on their borders.
The overset grid is also called as 
overlaid grid,
or composite overlapping grid,
or Chimera grid \citep{steger:1983}.
The validity and importance of the overset approach in
the aerodynamical calculations was
pointed out by \citet{steger:1982}.
Since then
this method is widely used in this field.
It is now one of the most important grid techniques
in the computational aerodynamics;
for example, 
whole aircraft with wing and store \citep{meakin:1992},
tiltrotor aircraft \citep{meakin:1993},
Boeing~747 \citep{cao:1998,rogers:1998},
Space Shuttle \citep{buning:1988},
helicopter \citep{duque:1996},
and others.

In the computational geosciences,
the idea of the overset grid approach appeared 
rather early.
\textit{Phillips} 
proposed a kind of composite grid
in 1950's to solve partial differential equations
on a hemisphere,
in which the high latitude region of 
the latitude-longitude grid is
``capped'' by another 
grid system that is constructed by a 
stereographic projection to a plane 
on the north pole \citep{phillips:1957,phillips:1959,browning:1989}.
After a long intermission,
the overset grid method seems to attract growing interest
in geoscience these days.
The ``cubed sphere'' \citep{ronchi:1996} 
is an overset grid that covers a spherical 
surface with six component grids that
correspond to six faces of a sphere.
The ``cubed sphere'' is recently applied to the 
mantle convection simulation \citep{hernlund:2003}.
In the atmospheric research,
other kind of spherical overset grid is used
in a global circulation model
\citep{dudhia:2002},
in which
the spherical surface is covered
by two component 
grids---improved stereographic projection grids---in
northern and souther hemispheres that
overlap in the equator.
A successful test of 100-day integration of global circulation
is demonstrated with this overset grid.

The overset grid proposed
in this paper 
is named ``Yin-Yang grid'' after the symbol
for yin and yang of Chinese philosophy of complementarity.
The Yin-Yang grid is composed of two identical and complemental 
component grids.
Compared with other spherical overset grids,
the Yin-Yang grid is simple in its geometry
and metric tensors.
A remarkable feature of this overset grid is that
the two identical component grids are combined in a complemental way
with a special symmetry.

\section{Basic Yin-Yang grid}
\begin{figure}[t]
 \begin{center}
  \includegraphics[width=0.75\textwidth]{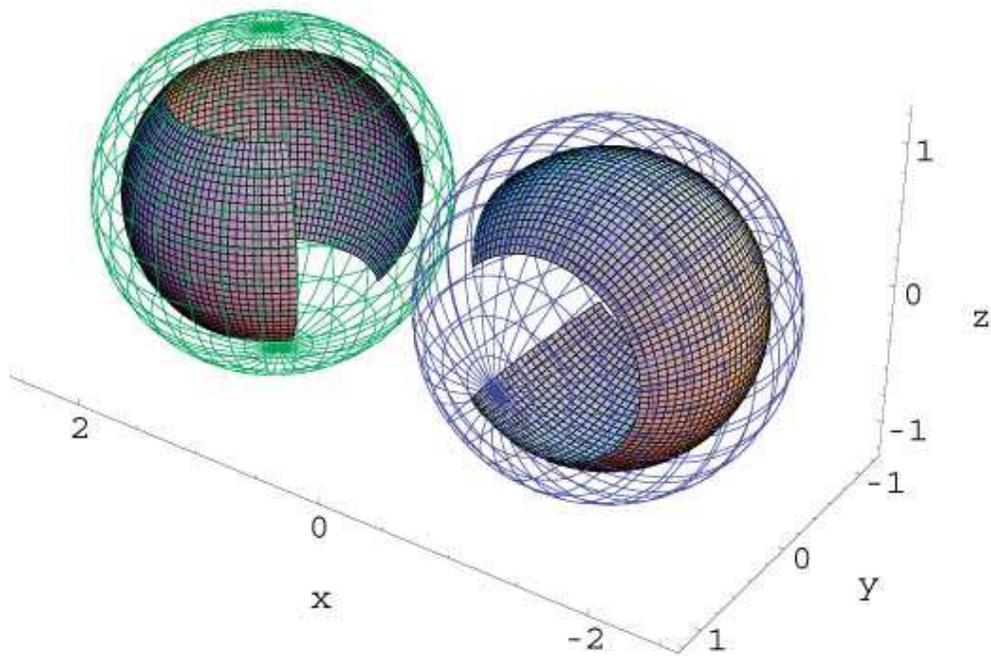}
 \end{center}
 \caption{The component grid of the Yin-Yang grid
is a part of the latitude-longitude grid.
The axes of two spherical polar coordinates
for Yin grid (green mesh) 
and Yang grid (blue mesh) are perpendicular.
}
 \label{fig:basicYinYangSphericalCoords}
\end{figure}

The Yin-Yang grid in its most basic shape
is shown in Fig.~\ref{fig:basicYinYang}.
It has two component grids that are geometrically identical
(exactly the same shape and size);
see Fig.~\ref{fig:basicYinYang}(a).
We call the two component grids ``Yin grid'' (or \textit{n}-grid) 
and ``Yang grid'' (or \textit{e}-grid).
They are combined to cover a spherical surface with partial overlap
on their borders as shown in Fig.~\ref{fig:basicYinYang}(b).
Each component grid is in fact
a part of the latitude-longitude grid:
A component grid, say Yin grid, 
is defined in the spherical polar coordinates by
\begin{equation}
  (\pi/4 - \delta \le \theta \le 3\pi/4 + \delta) 
       \cap 
  (-3\pi/4 - \delta \le \phi \le 3\pi/4 + \delta),	\label{eq:a00}
\end{equation}
where $\delta$ is a small buffer,
which is proportional to grid spacing, 
required for minimum overlap in the overset methodology
(see Fig.~\ref{fig:basicYinYang}(b)).
In the limit of infinitesimal grid ($\delta\rightarrow 0$),
the area of the above part of the sphere with unit radius is
given by
$\int_{\pi/4}^{3\pi/4} \sin \theta d\theta \int_{-3\pi/4}^{3\pi/4} d\phi 
= 3\pi/\sqrt{2} \sim 2.12 \pi$, i.e., roughly a half of the
whole spherical surface ($2\pi$).
Another component grid, Yang grid, is defined by the same
rule of eq.~(\ref{eq:a00}) but in different spherical coordinates
that is perpendicular to the original one;
see the green- and blue-colored spherical mesh in
Fig.~\ref{fig:basicYinYangSphericalCoords}.
The axis of the Yang grid's coordinates
(blue mesh in Fig.~\ref{fig:basicYinYangSphericalCoords}),
is located in a equator of the Yin grid's coordinates
(green mesh in Fig.~\ref{fig:basicYinYangSphericalCoords}).
The relation between Yin coordinates and Yang coordinates
is denoted in the Cartesian coordinates by
\begin{equation}
  (x^e, y^e, z^e) = (-x^n, z^n, y^n),		\label{eq:a04b}
\end{equation}
where $(x^n,y^n,z^n)$ is Yin's Cartesian coordinates
and $(x^e,y^e,z^e)$ is Yang's.
In a matrix form,
\begin{equation}
  \left(
    \begin{array}{c}
      x^e \\
      y^e \\
      z^e
    \end{array}
  \right)
  = 
  M 
  \left(
    \begin{array}{c}
      x^n \\
      z^n \\
      y^n
    \end{array} 
  \right),                              \label{eq:a05}
\end{equation}
where
\begin{equation}			
  M = \left(
        \begin{array}{ccc}
	  -1 & 0 & 0 \\
	   0 & 0 & 1 \\
	   0 & 1 & 0
	\end{array}
      \right).				\label{eq:a07}
\end{equation}
Note that
\begin{equation}
  M^{-1} = M,                           \label{eq:a06}
\end{equation}
which indicates that the transformations between Yin and Yang
coordinates are symmetric. 
This is a reflex of the complemental relation between Yin and Yang.

In the spherical coordinates, eq.~(\ref{eq:a04b}) reads
\begin{eqnarray}
     r^e & = & r^n, \\
      \sin\theta^e \cos\phi^e & = &  -\sin\theta^n \cos\phi^n,\label{eq:a05c}\\
      \sin\theta^e \sin\phi^e & = &   \cos\theta^n,\\
      \cos\theta^e & = &   \sin\theta^n \sin\phi^n,   \label{eq:a05b}
\end{eqnarray}
where $(r^n,\theta^n,\phi^n)$,
and $(r^e,\theta^e,\phi^e)$ are 
the coordinates of Yin and Yang, respectively.
The idea of two perpendicular spherical coordinates is used 
in the global ocean simulation \citep{eby:1994}
to avoid the grid convergence in the Arctic,
however, the second spherical coordinates
is used in a sort of auxiliary way for the main
(usual) spherical polar coordinates in their method.
On the other hand, we make the best use of the
symmetry between two coordinates.

For spatial discretization,
we define mesh point at 
$j$-th colatitude $\theta_j^\ell$
and $k$-th longitude $\phi_k^\ell$
on Yin grid (for $\ell=n$) and on Yang grid (for $\ell=e$) as
\begin{eqnarray}
  \theta^\ell_j &=& 
     \theta_{\hbox{min}}+j\, \Delta\theta, \ \ (j=0, N_\theta-1),\\
  \phi^\ell_k &=& 
     \phi_{\hbox{min}}+k\, \Delta\phi, \ \ (k=0, N_\phi-1),
\end{eqnarray}
with
\begin{eqnarray}
 \Delta\theta&=&(\theta_{\hbox{max}}-\theta_{\hbox{min}})/(N_\theta-1),\\
 \Delta\phi&=&(\phi_{\hbox{max}}-\phi_{\hbox{min}})/(N_\phi-1),
\end{eqnarray}
where the grid distribution ranges from
$\theta_{\hbox{min}}=\pi/4-\delta$
to $\theta_{\hbox{max}}=3\pi/4+\delta$
in colatitude, and from
$\phi_{\hbox{min}}=-3\pi/4-\delta$
to $\phi_{\hbox{max}}=3\pi/4+\delta$
in longitude.
We set $\Delta\theta=\Delta\phi=2 \delta$
in Fig.~\ref{fig:basicYinYang}, as an example.

An important feature of the Yin-Yang grid as a spherical overset grid
is that the two component grids are identical and
their geometrical positions are complemental.
This enables us to make concise programs:
Suppose a grid point $(\theta_j^n,\phi_k^n)$
on Yin grid's horizontal border
at index position $(j,k)$ (e.g., $j=1$).
Its value should be determined by an interpolation from
its neighbor points, or stencils, of Yang grid with
interpolation coefficients that are 
determined by relative position of 
$(\theta_j^n,\phi_k^n)$
in the stencils.
Note that exactly the same interpolation coefficients and 
relative stencils are used to set the value of corresponding grid point 
$(\theta_j^e,\phi_k^e)$
at $(j,k)$ of Yang's border,
since the geometrical
relations between Yin grid and Yang grid are symmetric.
In other words, we can make use of one interpolation routine for two times
(for Yin grid and for Yang grid)
to set the horizontal boundary conditions.
Note also that the metric tensors at a bulk
grid point at $(j,k)$ of Yin grid is a function
of its position $(\theta_j^n,\phi_k^n)$ in Yin's coordinates,
and the metric tensors at corresponding point 
$(\theta_j^e,\phi_k^e)$ in Yang grid
are exactly the same.
Therefore we can call one subroutine of fluid solver and others
for two times for Yin grid and Yang grid.

Another advantage of the Yin-Yang grid
resides in the fact that the component grid is 
nothing but the (part of ) latitude-longitude grid.
We can directly deal with the equations to be solved with 
the vector form in the usual spherical polar coordinates,
$\{v_r, v_\theta, v_\phi\}$.
The analytical form of metric tensors are familiar
in the spherical coordinates.
We can directly code
the basic equations 
in the program as they are formulated in the
spherical coordinates.
We can make use of various resources of
mathematical formulas, program libraries, and tools 
that have been developed in the spherical polar coordinates.

To conclude this section,
we point out that
the construction of three-dimensional
Yin-Yang grid
for spherical shell geometry is straightforward,
by piling up the basic (two-dimensional) 
Yin-Yang grids in radial direction.
See Fig.~\ref{fig:basicYinYang3D}.

\section{Vector transformation formula between Yin and Yang grids}
Following the general overset methodology
\citep[e.g.,][]{chesshire:1990},
interpolations are applied on the boundary of each component grid
to set the boundary values, or internal boundary condition.
When one deals with scalar variables,
the interpolation is simple.
For vector fields, a care is needed
for vector components, since expressions of a vector in
the Yin's spherical coordinates,
$\{v^n_r,v^n_\theta,v^n_\phi\}$,
and in the Yang's coordinates,
$\{v^e_r,v^e_\theta,v^e_\phi\}$, are different.

\begin{figure}[t]   
 \begin{center}
  \includegraphics[width=0.8\textwidth]{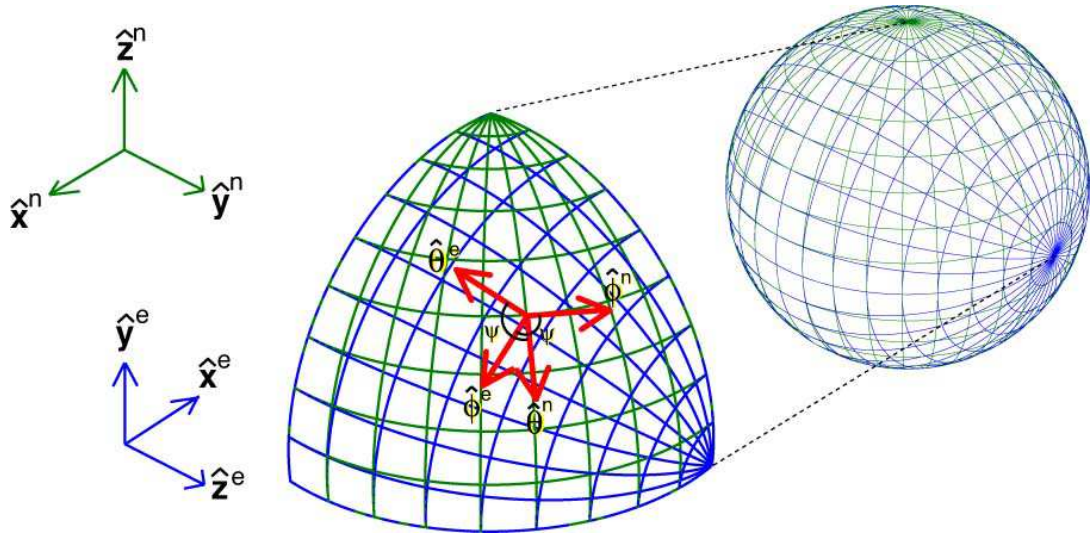}
 \end{center}
 \caption{Unit horizontal vectors of Yin and Yang coordinates.
They are mapped one another by 
a local rotation transform with angle $\psi$.
}
 \label{fig:sphericalNormalVectors}
\end{figure}

Because the Yin-Yang transformation denoted by eq.~(\ref{eq:a04b})
is a rotation about the origin ($r=0$),
the radial component of the vector is invariant
($v_r^n=v_r^e$), and 
horizontal components are
mapped by local rotation transforms,
as shown in Fig.~\ref{fig:sphericalNormalVectors},
where the rotation angle $\psi$ is a function of
latitude and longitude;
\begin{equation}
  \left(
    \begin{array}{c}
      v_r^e\\
      v_\theta^e \\
      v_\phi^e
    \end{array}
  \right)
  = 
  \left(
   \begin{array}{ccc}
    1 & 0          & 0 \\
    0 & \cos \psi  & -\sin \psi\\
    0 & \sin \psi  &  \cos \psi
   \end{array}
  \right)
  \left(
    \begin{array}{c}
      v_r^n      \\
      v_\theta^n \\
      v_\phi^n
    \end{array} 
  \right).                              \label{eq:a08}
\end{equation}
To find the expression of $\psi$,
we consider 
unit vectors in $\theta$ and $\phi$ directions
on the Yin and Yang coordinates.
From Fig.~\ref{fig:sphericalNormalVectors}, 
we see 
\begin{eqnarray}
    \cos\psi & = & \hat{\phi}^n \cdot \hat{\phi}^e ,	\label{eq:a12}\\
    \sin\psi & = & -\hat{\phi}^n \cdot \hat{\theta}^e,	\label{eq:a13}
\end{eqnarray}
where $\hat{\theta}^\ell$ and $\hat{\phi}^\ell$ are unit vectors 
in $\theta$ and
$\phi$ directions in the component grid $\ell$,
with $\ell=n$ for Yin grid, and $\ell=e$ for Yang grid.
The unit vectors \{$\hat{x}^\ell, \hat{y}^\ell$, $\hat{z}^\ell$\} in the Cartesian
coordinates are related to \{$\hat{\theta}^\ell$, $\hat{\phi}^\ell$\} by
\begin{eqnarray}
    \hat{\phi}^e & = & 
       -\sin \phi^e \, \hat{x}^e + \cos \phi^e \, \hat{y}^e,	\label{eq:a14}\\
    \hat{\phi}^n & = & 
       -\sin \phi^n \, \hat{x}^n + \cos \phi^n \, \hat{y}^n 	\nonumber\\     
                 & = &
        \sin \phi^n \, \hat{x}^e + \cos \phi^n \, \hat{z}^e,	\label{eq:a15} \\
    \hat{\theta}^e & = & \cos \theta^e  \cos \phi^e \, \hat{x}^e
                 +   \cos \theta^e  \sin \phi^e \, \hat{y}^e
                 -   \sin \theta^e  \hat{z}^e.			\label{eq:a16}
\end{eqnarray}
Substituting eqs.~(\ref{eq:a14}) and~(\ref{eq:a15}) into~(\ref{eq:a12}),
we get  
\begin{equation}
  \cos\psi = - \sin\phi^e\, \sin\phi^n.				\label{eq:a17}
\end{equation}
Substituting eqs.~(\ref{eq:a15}) and~(\ref{eq:a16}) into~(\ref{eq:a13}),
we get  
\begin{eqnarray}
  \sin\psi &=& \cos\theta^e \cos\phi^e \sin\phi^n - \sin\theta^e \cos\phi^n \nonumber\\
           &=& \frac{1}{\sin\theta^e \sin\theta^n}
              \, \left\{
                   \cos\theta^e\, (\sin\theta^e \cos\phi^e)\,
                                 (\sin\theta^n \sin\phi^n)
                       - \sin^2\theta^e\, (\sin\theta^n \cos\phi^n)
                 \right\}   \nonumber\\
           &=& -\frac{\cos\phi^e}{\sin\theta^n}	\nonumber\\
           &=&  \frac{\cos\phi^n}{\sin\theta^e}.		\label{eq:a18}
\end{eqnarray}
Here we have used eqs.~(\ref{eq:a05c})--(\ref{eq:a05b}).

From eqs.~(\ref{eq:a17}), (\ref{eq:a18}),
(\ref{eq:a12}), (\ref{eq:a13}) and~(\ref{eq:a08}), we obtain
the transformation formula of the vector components
$(v_r, v_\theta, v_\phi)$ by
\begin{equation}
  \left( 
    \begin{array}{c}
      v_r^e \\
      v_\theta^e\\
      v_\phi^e
    \end{array}
  \right) 
  =
  P
  \left(
    \begin{array}{c}
      v_r^n \\
      v_\theta^n \\
      v_\phi^n
    \end{array}
  \right),
\end{equation}
with the transformation matrix
\begin{equation}
  P =
  \left(
    \begin{array}{ccc}
      1 & 0 & 0 \\
      0 & -\sin{\phi^e} \sin{\phi^n} & - \cos{\phi^n} / \sin{\theta^e}\\
      0 & \cos{\phi^n} / \sin{\theta^e} & -\sin{\phi^e} \sin{\phi^n} 
    \end{array}
  \right).
\end{equation}
Since 
Yin and Yang coordinates are symmetric, the inverse transformation
from Yang into Yin is given by the interchange
of the suffixes:
\begin{equation}
  P^{-1} = 
  \left(
    \begin{array}{ccc}
      1 & 0 & 0 \\
      0 & -\sin{\phi^n} \sin{\phi^e} & - \cos{\phi^e} / \sin{\theta^n}\\
      0 & \cos{\phi^e} / \sin{\theta^n} & -\sin{\phi^n} \sin{\phi^e} 
    \end{array}
  \right).
\end{equation}
Note also that
\begin{equation}
  P^2=1,
\end{equation}
which
indicates the complemental relation between Yin and Yang coordinates.

\begin{figure}[t]   
 \begin{center}
  \includegraphics[width=0.7\textwidth]{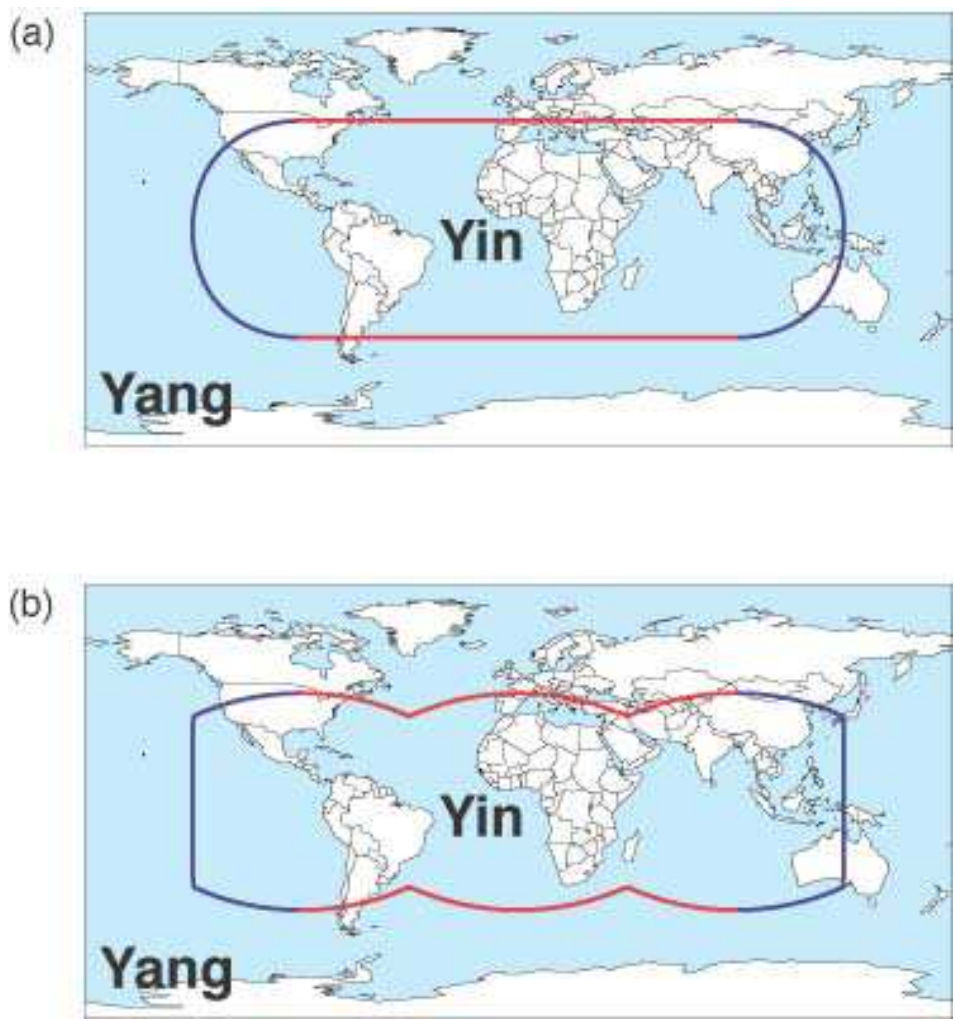}
 \end{center}
 \caption{Curves that divide a spherical surface into two identical areas.
If one cuts along the blue-red curve of the panel (a) or (b),
the spherical surface is divided into two parts (denoted by 
Yin and Yang in the pictures)
that are exactly
the same shape and size.
The blue part of the curve and red part of the curve 
are in the complemental relation;
The blue curve of Yin is red curve of Yang, and vice versa.
}
 \label{fig:boundaryCurveWithMap}
\end{figure}

\begin{figure}[t]   
 \begin{center}
  \includegraphics[width=0.99\textwidth]{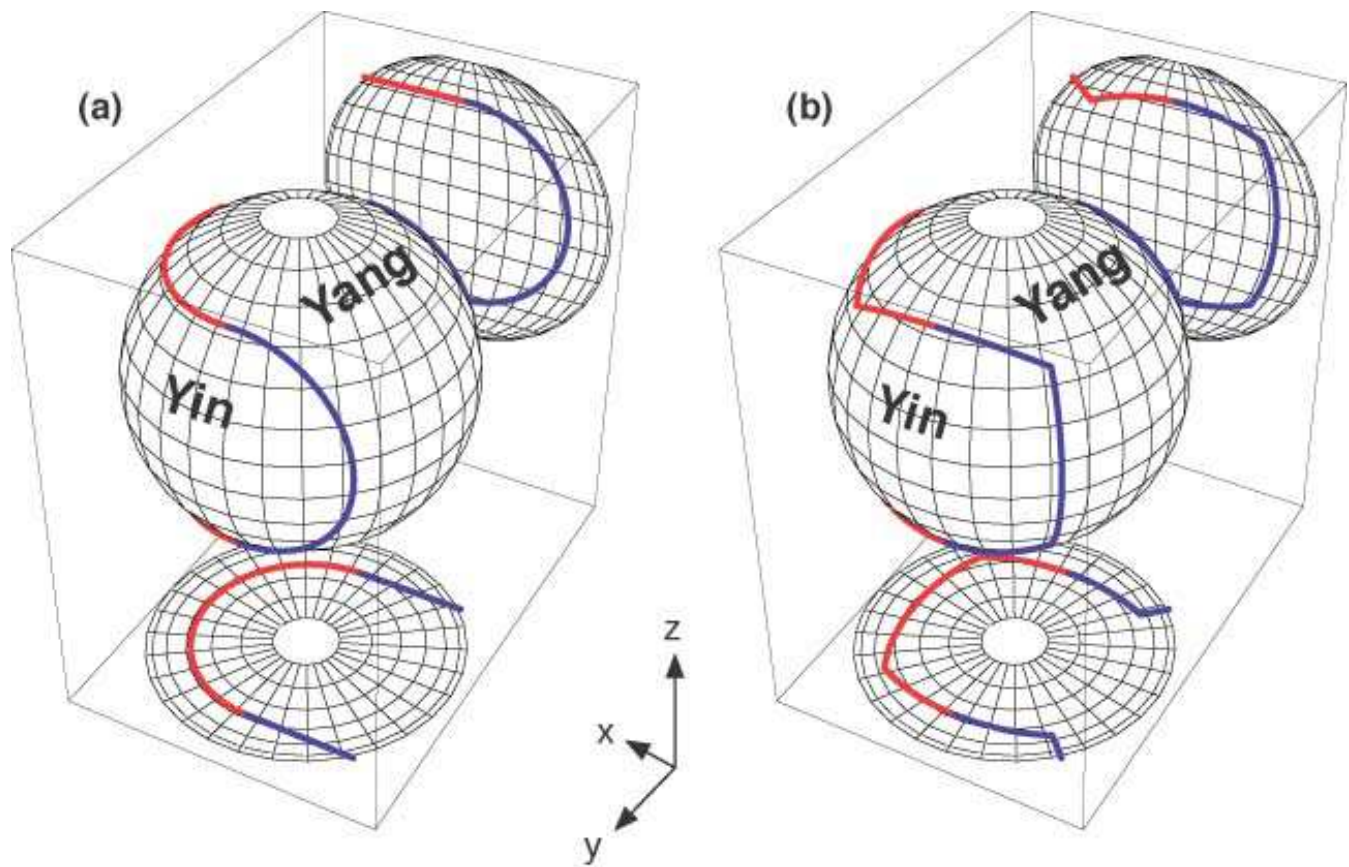}
 \end{center}
 \caption{Curves that divide a spherical surface into two identical areas.
These are corresponding three-dimensional views of 
the red-blue curves in (a) and (b)
of Fig.~\ref{fig:boundaryCurveWithMap}.
}
 \label{fig:boundaryCurve3D}
\end{figure}

When we see the component grid of the basic Yin-Yang grid
shown in Fig.~\ref{fig:basicYinYang}
in the Mercator projection, it is a rectangle;
the four corners
intrude most into the other component grid
(see Fig.~\ref{fig:basicYinYang}(b)).
Even if the grid mesh is taken to be infinitesimal,
i.e., $\Delta\theta=\Delta\phi\rightarrow 0$ and $\delta\rightarrow 0$,
the overlapping area has still non-zero ratio of about $6\%$;
$(3/\sqrt{2}\pi-2\pi)/2\pi\sim 0.061$.
This overlapped area can be minimized by
modifying the component grid's shape from the 
rectangle.
It is obvious that 
a Yin-Yang grid with minimum overlap region
can be constructed by a division,
or dissection, 
with a closed curve 
on a sphere that cuts the sphere into two identical parts.
There are infinite number of such dissections of a sphere.
Fig.~\ref{fig:boundaryCurveWithMap} shows two examples among them.
When we cut along the curve
that is colored with red and blue in Fig.~\ref{fig:boundaryCurveWithMap}(a)
or (b),
we get two separated parts of the spherical surface that
are identical.
Although, it is not apparent that
the two parts separated by the blue-red curve in each
panel of Fig.~\ref{fig:boundaryCurveWithMap}
are identical from this figure,
the corresponding 
three-dimensional view (Fig.~\ref{fig:boundaryCurve3D})
would show more convincingly.
The cutoff curve of Fig.~\ref{fig:boundaryCurve3D}(a) 
reminds us a baseball,
while the cutoff curve of Fig.~\ref{fig:boundaryCurve3D}(b) 
resembles a cube.

\begin{figure}[t]   
 \begin{center}
  \includegraphics[height=0.95\textheight]{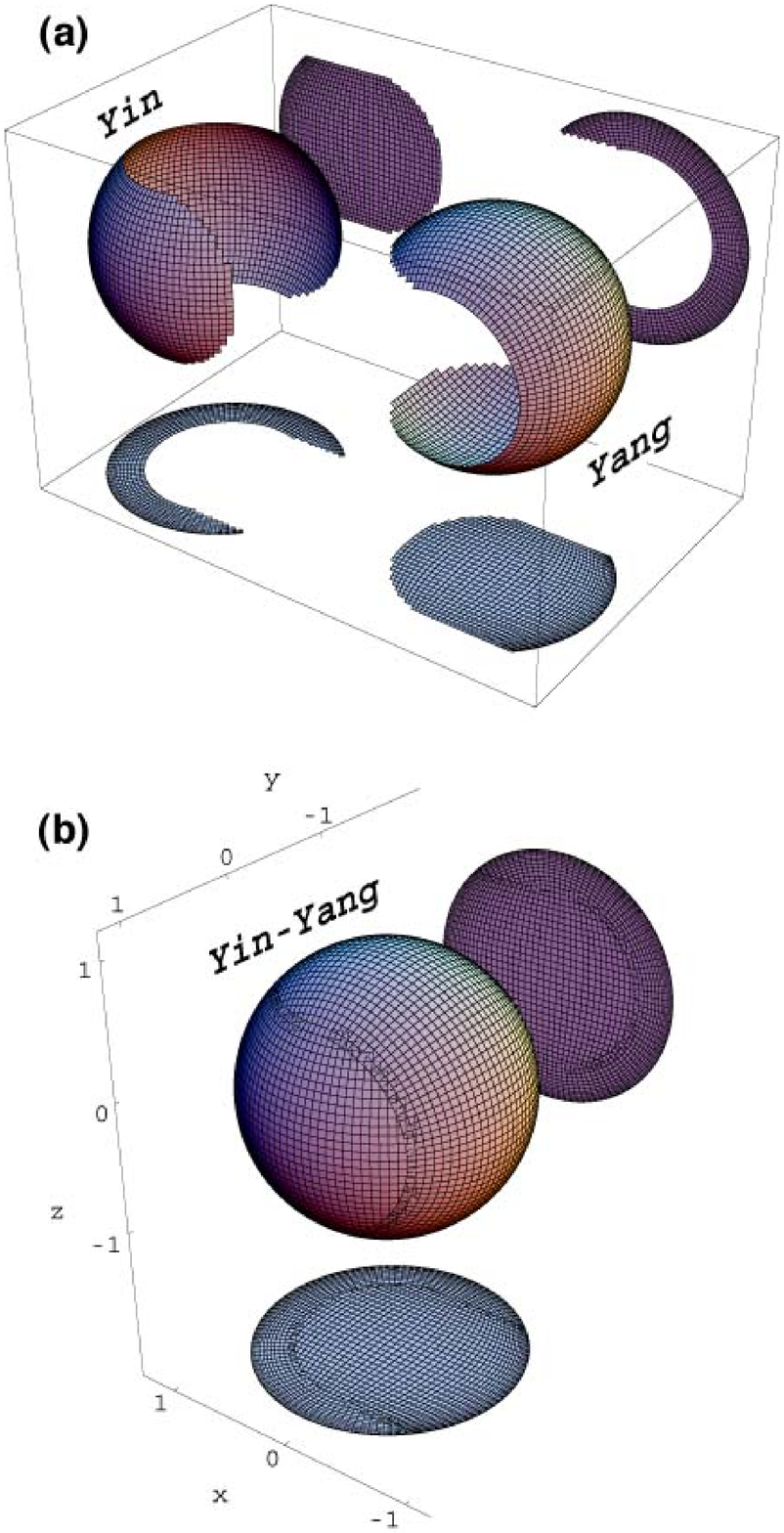}
 \end{center}
 \caption{A Yin-Yang grid with minimum overlap
that has the baseball-like border curve
between Yin and Yang grids.
Corresponding spherical dissection 
is Figs.~\ref{fig:boundaryCurveWithMap}(a)
and~\ref{fig:boundaryCurve3D}(a).}
 \label{fig:stadiumYinYang}
\end{figure}

\begin{figure}[t]   
 \begin{center}
  \includegraphics[height=0.95\textheight]{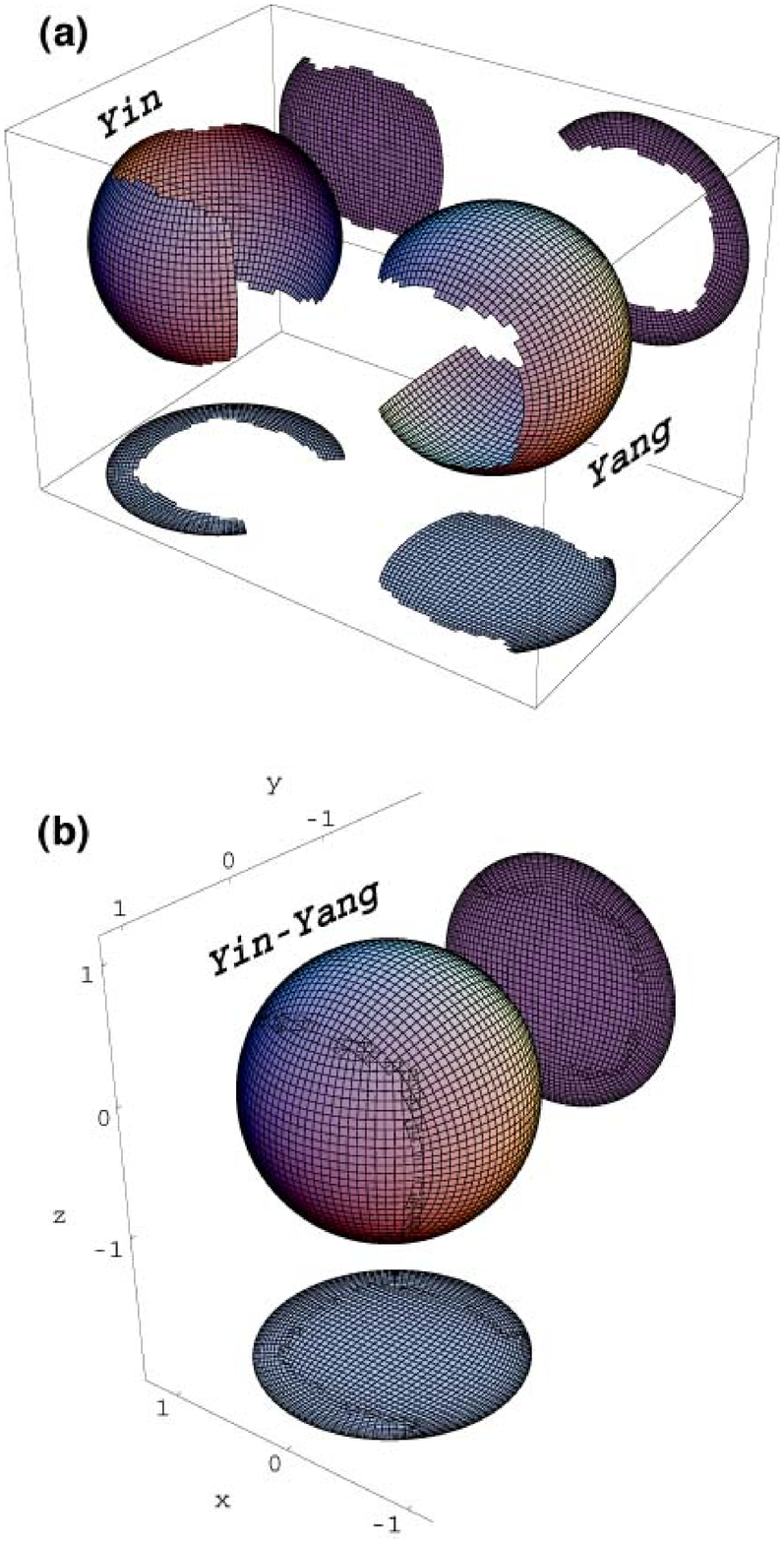}
 \end{center}
 \caption{Another Yin-Yang grid with minimum overlap.
The border curve between Yin and Yang grids
is cube-like.
Corresponding spherical dissection 
is Figs.~\ref{fig:boundaryCurveWithMap}(b)
and~\ref{fig:boundaryCurve3D}(b).
}
 \label{fig:diceYinYang}
\end{figure}

Based on these spherical dissections, 
we can construct spherical overset grids with two
identical component grids that has minimum overlapping area;
Fig.~\ref{fig:stadiumYinYang} shows
a Yin-Yang grid that corresponds to
the baseball type dissection of a sphere (panels labeled~(a) in
Figs.~\ref{fig:boundaryCurveWithMap} and~\ref{fig:boundaryCurve3D}).
Fig.~\ref{fig:diceYinYang} is for the cube type 
dissection 
(panels~(b) of Figs.~\ref{fig:boundaryCurveWithMap} 
and~\ref{fig:boundaryCurve3D}).
When minimizing the computational cost is strongly required,
the Yin-Yang grid of the baseball type
(Fig.~\ref{fig:stadiumYinYang})
or cube type (Fig.~\ref{fig:diceYinYang})
would be worth trying.

However, the non-rectangle geometries of
the component grid of 
Fig.~\ref{fig:stadiumYinYang}
or
Fig.~\ref{fig:diceYinYang}
imply that special cares should be taken to mask some grid points.
The number of the mask is the same for both the Yin-Yang grids
of Figs.~\ref{fig:stadiumYinYang} and~\ref{fig:diceYinYang},
since the non-masked area of a component grid is just
a half of the spherical surface ($2\pi$)
in the limit of the negligibly small overlap area.

\section{Summary}
For numerical simulations of the Earth's interior,
we have developed a new spherical grid based on 
the overset grid methodology.
Our motivation is to devise an spherical grid system
that is suitable for finite difference scheme on
massively parallel vector supercomputers.
The spherical overset grid proposed in this paper,
named Yin-Yang grid, is composed of two component grids.
They have the same shape and size and combined to cover
a spherical surface with partial overlap on their borders.
Each component grid is nothing but low latitude region
of the usual latitude-longitude grid;
it is $90^\circ$ about the equator and
$270^\circ$ in the longitude.
Therefore the grid spacing is quasi-uniform
and the metric tensors are simple and analytically known.
One can directly apply 
mathematical and numerical resources that have been written
in the spherical polar coordinates or latitude-longitude grid system.
Since the two component grids are identical and
combined in a complemental way,
various routines for solvers and interpolation
can be recycled for two times for each component grid
at every simulation time step.

\begin{figure}[t]
 \begin{center}
  \includegraphics[height=0.95\textheight]{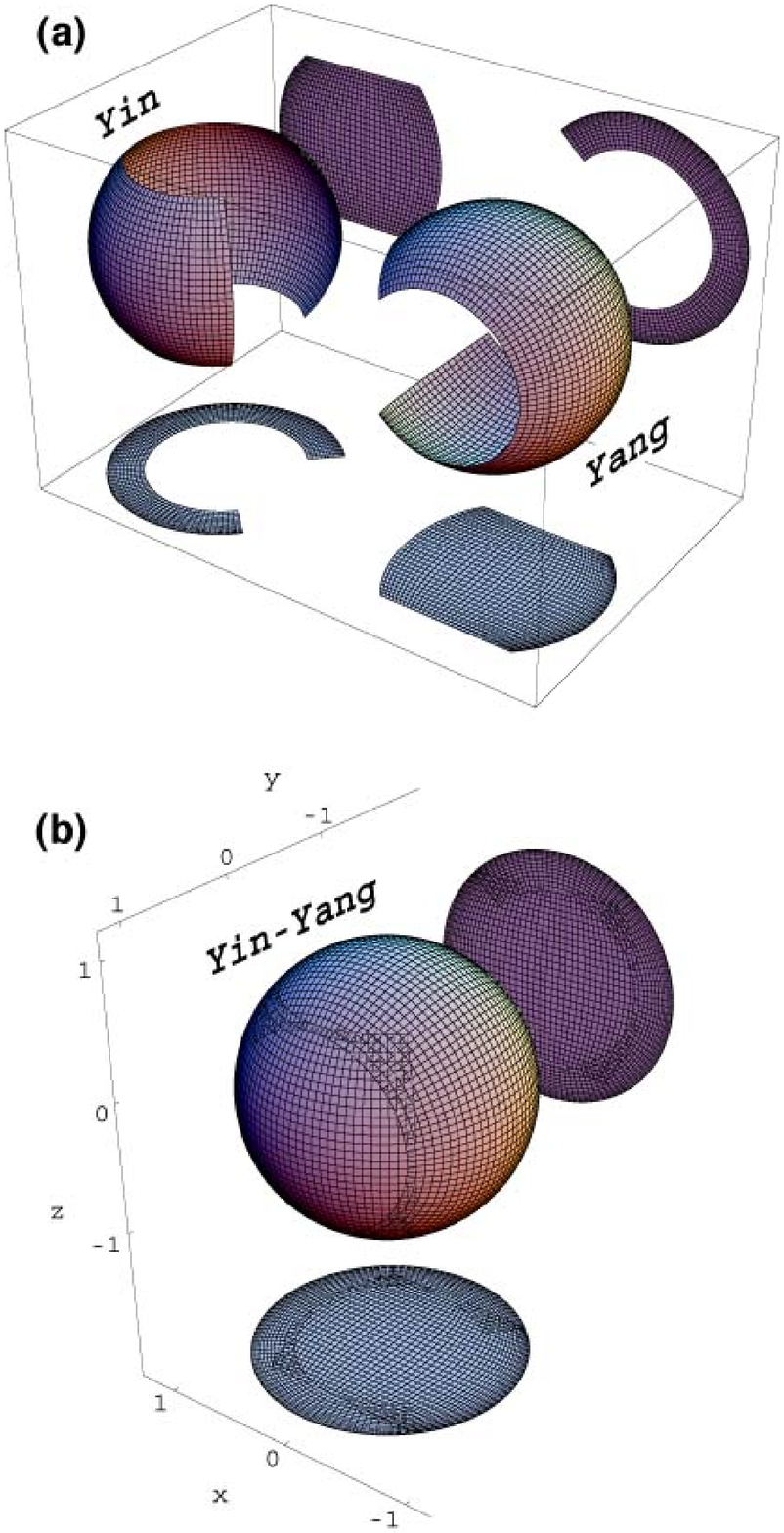}
 \end{center}
 \caption{Basic Yin-Yang grid.
(a) It is a spherical overset grid 
composed of two identical component grids,
named Yin and Yang.
(b) The Yin grid and Yang grid 
are combined to
cover a spherical surface with partial overlap.
}
 \label{fig:basicYinYang}
\end{figure}

\begin{figure}[t]
 \begin{center}
  \includegraphics[width=0.7\textwidth]{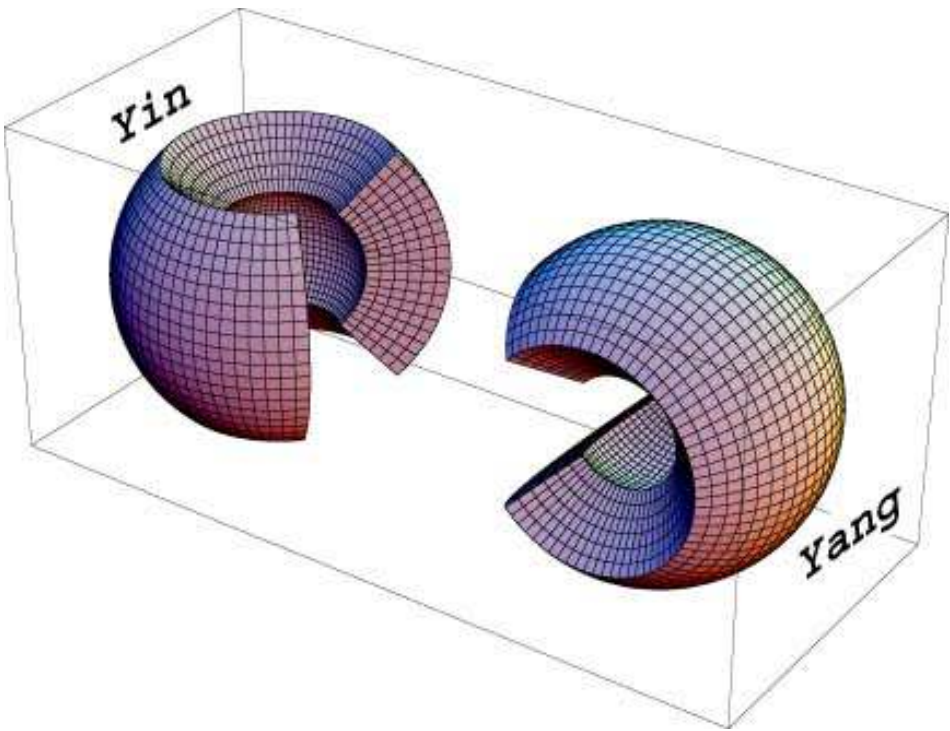}
 \end{center}
 \caption{Three-dimensional Yin-Yang grid for spherical shell geometry.
This is constructed by piling up the basic Yin-Yang grid
shown in Fig.~\ref{fig:basicYinYang} in the radial direction.
 \label{fig:basicYinYang3D}
}
\end{figure}

We have developed finite difference codes of
the mantle convection and 
dynamo simulation using 
the basic Yin-Yang grid for spherical shell
geometry (see Figs.~\ref{fig:basicYinYang}
and \ref{fig:basicYinYang3D}).
We have confirmed that the Yin-Yang grid is successfully
applied to both cases.
The mantle convection code is newly developed from scratch.
Details of the code and simulation results
are reported in other paper \citep{yoshida:2004};
we solved the time development of thermal convection motion
in a spherical shell of a Boussinesq fluid
with infinite Prandtl number
for uniform and variable viscosity cases.
We have performed standard benchmark tests of the mantle convection
\citep{richards:2001},
and confirmed that the results of
our Yin-Yang mantle convection code successfully reproduced 
previously published results.
The numerical values of Nusselt number
and the mean velocity coincides with
other benchmark values within a few percent or even 
better \citep{yoshida:2004}.
We have also applied the Yin-Yang grid to the geodynamo simulation code.
The magnetohydrodynamic (MHD) equations
with finite viscosity, thermal diffusivity, and electrical conductivity
are solved.
The Yin-Yang geodynamo code 
has been converted from our previous geodynamo code which was based
on the latitude-longitude grid.
We found that the code conversion was straightforward and
rather easy since the base grid is common.
We could reproduce our previous (latitude-longitude grid based)
results of geodynamo simulation
by our newly developed Yin-Yang geodynamo code
with shorter calculation time.
The details of the code will be reported in other paper.

The Yin-Yang grid is 
suitable for parallel programming.
Since the number of the component grid is two,
we are naturally lead to make parallel programs with
domain decomposition of even number:
We first decompose whole computational region into 
two---Yin component and Yang component---then 
apply further domain decomposition in each component.

\begin{figure}[t]   
 \begin{center}
  \includegraphics[width=0.9\textwidth]{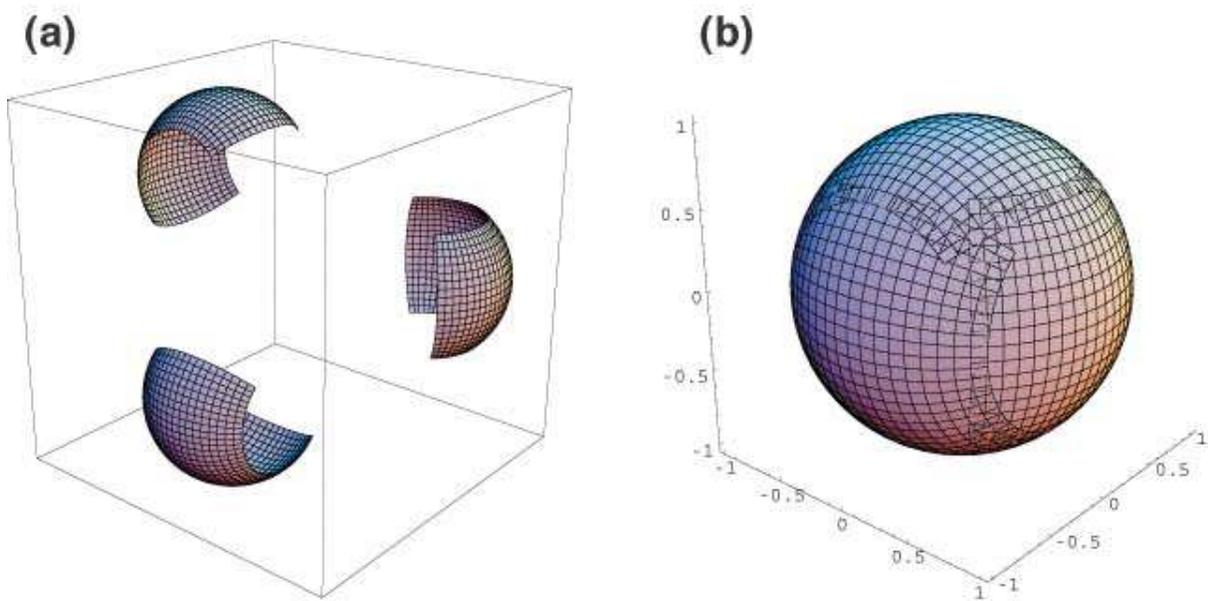}
 \end{center}
 \caption{Another possible spherical overset grid
that is composed of three identical component grids.}
 \label{fig:chimera3elements}
\end{figure}

Finally, we point out another possible spherical overset 
grid that has an odd number of component grids.
Fig.~\ref{fig:chimera3elements} shows a spherical overset grid
that consists of three identical component grids.
In this case, the component grid is defined as a part
(about $1/3$) of the spherical surface by
$(\pi/4\le\theta\le 3\pi/4)\cap
(-\pi/2\le\phi\le \pi/2)$.
This grid could be
effective when the processor number is multiple of three.

\begin{acknowledgments}
We would like to acknowledge helpful discussion
with: Masanori Kameyama, Kenji Komine, Hideaki Miura, Keiko Takahashi,
and Masaki Yoshida.
The development and benchmark test of
the mantle convection code using the Yin-Yang grid
was done by Masaki Yoshida.
All simulations were performed by Earth Simulator,
Japan Agency for Marine-Earth Science and Technology.
\end{acknowledgments}

\newpage

%
%

\end{document}